%% file: main.tex
\documentclass[manuscript,times]{aastex62}
\usepackage{glossaries}

\usepackage{soul}

\hypersetup{breaklinks=true}
\def\UrlBreaks{\do\/\do-} 

\graphicspath{{./}{figures/}}

\shorttitle{Roman Coronagraph Exozodi Sensitivity}
\shortauthors{Douglas et al.}

\input{acronyms}

\newacronym{OS}{OS}{observing scenario}
\newacronym{LOWFSC}{LOWFSC}{low-order wavefront sensing and control}
\newacronym{HOWFSC}{LOWFSC}{high-order wavefront sensing and control}
\begin{document}
\title{Sensitivity of the Roman Coronagraph Instrument to Exozodiacal Dust}

\correspondingauthor{Ewan S. Douglas}
\email{douglase@arizona.edu}
\author[0000-0002-0813-4308]{Ewan S Douglas}
\affiliation{Department of Astronomy and Steward Observatory, University of Arizona, 933 N. Cherry Ave, Tucson, AZ 85721, USA}

\author[0000-0002-1783-8817]{John Debes}
\affiliation{Space Telescope Science Institute, 3700 San Martin Drive, Baltimore, MD 21218, USA}

\author[0000-0003-4205-4800]{Bertrand Mennesson}
\affiliation{Jet Propulsion Laboratory, California Institute of Technology, 4800 Oak Grove Drive, Pasadena, CA 91109, USA}

\author[0000-0002-2096-4187]{Bijan Nemati}
\affiliation{Center for Applied Optics, The University of Alabama in Huntsville,
301 Sparkman Drive, Huntsville, AL 35899}

\author[0000-0001-5082-7442]{Jaren Ashcraft}
\affiliation{Wyant College of Optical Sciences, University of Arizona, Tucson, AZ 85721, USA}

\author[0000-0003-1698-9696]{Bin Ren}  

\affiliation{Department of Astronomy, California Institute of Technology, MC 249-17, 1200 East California Boulevard, Pasadena, CA 91125, USA}

\author[0000-0002-2805-7338]{Karl Stapelfeldt}
\affiliation{Jet Propulsion Laboratory, California Institute of Technology, 4800 Oak Grove Drive, Pasadena, CA 91109, USA}

\author[0000-0002-8711-7206]{Dmitry Savransky}
\affil{Sibley School of Mechanical and Aerospace Engineering, Cornell University, Ithaca, NY 14853, USA}
\affil{Carl Sagan Institute, Cornell University, Ithaca, NY 14853, USA}
\author[0000-0002-8507-1304]{Nikole K. Lewis}
\affil{Department of Astronomy, Cornell University, 122 Sciences Drive, Ithaca, NY 14853, USA}
\affil{Carl Sagan Institute, Cornell University, Ithaca, NY 14853, USA}
\author[0000-0003-1212-7538]{Bruce Macintosh}
\affil{Kavli Institute for Particle Astrophysics and Cosmology, Department of Physics, Stanford University, Stanford, CA, 94305, USA}

\newacronym{resel}{resel}{resolution element}
\newacronym{HOSTS}{HOSTS}{Hunt for Observable Signatures of Terrestrial Systems}
\newacronym{HIP}{HIP}{Hipparchus}
\newacronym{HZ}{HZ}{habitable zone}
\newacronym{Roman}{Roman}{Nancy Grace Roman Space Telescope}
\newacronym{OWA}{OWA}{outer-working angle}
\newacronym{CIC}{CIC}{clock-induced charge}




\begin{abstract}
Exozodiacal dust, warm debris from comets and asteroids in and near the habitable zone of stellar systems, reveals the physical processes that shape planetary systems.
Scattered light from this dust is also a source of background flux which must be overcome by future missions to image Earthlike planets.
This study quantifies the sensitivity of the Nancy Grace Roman Space Telescope Coronagraph to light scattered by exozodi, the zodiacal dust around other stars.
Using a sample of 149 nearby stars, previously selected for optimum detection of habitable exoplanets by space observatories, we find the maximum number of exozodiacal disks with observable \textit{inner} habitable zone boundaries is {six} and the number of observable outer habitable boundaries is {74}.
One zodi was defined as the visible-light surface brightness of 22 $m_{\rm V}\ $arcsec$^{-2}$ around a solar-mass star, approximating the scattered light brightness in visible light at the Earth-equivalent insolation.
In the speckle limited case, where the signal-to-noise ratio is limited by speckle temporal stability rather than shot noise, the median $5\sigma$ sensitivity to habitable zone exozodi is {12}  zodi per resolution element.
This estimate is calculated at the inner-working angle of the coronagraph, for the current best estimate performance, neglecting margins on the uncertainty in instrument performance and including a post-processing speckle suppression factor.
For an log-norm distribution of exozodi levels with a median exozodi of 3$\times$ the solar zodi,
we find that the Roman Coronagraph would be able to make 5$\sigma$ detections of exozodiacal disks in scattered light from {13} systems with a 95\% confidence interval spanning {7-20} systems. 
This sensitivity allows Roman Coronagraph to complement ground-based measurements of exozodiacal thermal emission and constrain dust albedos.
Optimized post-processing and detection of extended sources in multiple resolution elements is expected to further improve this unprecedented sensitivity to light scattered  by exozodiacal dust.

\end{abstract}

\section{Introduction} \label{sec:intro}
Scattered light from warm (${\gtrsim}150$ K) circumstellar grains, zodiacal dust, is the brightest integrated visible-light source in the inner solar system \citep{gaidos_observational_1998,nesvorny_cometary_2010}; however, existing instrumentation is unable to directly image scattered light from this warm dust, primarily due to contamination from diffracted stellar light, which requires a state-of-the-art coronagraph to suppress.
The size distribution, morphology, and composition of this warm debris is driven by a variety of dynamical processes \citep{gustafson_physics_1994,stark_detectability_2008,kral_exozodiacal_2017} and evolutionary processes \citep{wyatt_evolution_2008}.
While astronomers have observed large cold debris disks far from stars, these analogs to the Kuiper (or Edgeworth-Kuiper) belt are generally outside the snowline and driven by slow collisional processes.
Conversely, little is known about extrasolar analogs to the solar system zodiacal dust belt.
This warm ``exozodi'' is populated by Poynting-Robertson drag, inwardly drifting debris from cometary and asteroid collisions \citep{gustafson_physics_1994,nesvorny_cometary_2010}.
Many larger debris particles orbit on bound orbits while smaller particles are more often ejected from the solar system by photon pressure or the solar wind \cite{szalay_near-sun_2020}.
Our understanding of the role these various forces play in extrasolar systems is severely limited by a lack of  observations to constrain albedo, dust size distributions, or morphology of the disks around other stars. 
For a more comprehensive review of debris disk architectures, properties, variability and observables, see \cite{hughes_debris_2018}.

Ground-based high-contrast instruments and extreme AO have observed cold debris disks at visible and NIR wavelengths (e.g., \citealp{rodigas_gray_2012,perrin_polarimetry_2015,rodigas_magao_2016,currie_subaruscexao_2017,schmid_sphere/zimpol_2018,duchene_gemini_2020,esposito_debris_2020} and many others) but lack the contrast and \gls{IWA} to detect  exozodiacal dust near stars.
Likewise, ALMA has observed continuum emission from protoplanetary systems and massive cold dust disks at millimeter wavelengths (e.g., \citealp{macgregor_alma_2016,andrews_disk_2018}) but lacks the sensitivity to resolve more rarefied exozodiacal dust.
Probing closer to stars, \gls{WISE} \citep{wright_wide-field_2010} all-sky survey data has been used to search for the \gls{IR} excess of warm dust (e.g., \citealp{patel_sensitive_2014} and references therein). 
However, due to the low spatial resolution of WISE, many of these detections are false positives  (see \citealp{silverberg_follow-up_2018,dennihy_word_2020}). 
High-spatial resolution direct imaging is less prone to confusion and
 \gls{HST} coronagraphs have been used to observe  circumstellar environments of several stars (e.g., \citealp{kalas_planetary_2005,krist_hubble_2012,schneider_probing_2014,ren_post-processing_2017,debes_pushing_2019}) but sensitivity is limited to orders of magnitude above solar system levels and limited to relatively large on-sky separations $>$0.5\farcs. 
 This limitation arises from a variety of coronagraph performance factors, including orbit-induced thermal variation and jitter (see \cite{krist_high-contrast_2004} and \cite{debes_pushing_2019-1}).

Since most reflected starlight and many biomarkers of interest  fall at visible and \gls{NIR} wavelengths \citep{schwieterman_exoplanet_2018}, the scattering of visible light by dust is the primary source of background flux that will limit detailed spectroscopic characterization of exoplanet atmospheres.
The telescope aperture which provides a significant number of Earthlike exoplanet detections (and spectra) is set, in part, by the background exozodiacal dust signal \citep{backman_exozodiacal_1998,defrere_nulling_2010,roberge_exozodiacal_2012,stark_direct_2016,turnbull_search_2012,stark_maximizing_2014}.
Precursor observations to help constrain the background flux arising from exozodiacal dust will also provide a sample of stellar systems to test theories of dust transport and morphology (e.g., \citealp{wyatt_insignificance_2005,stark_detectability_2008,hughes_debris_2018}).

\gls{IR} nulling surveys have proven a fruitful way to understand circumstellar dust statistically.
 By assuming solar-like disk properties and morphology \citep{kelsall_cobe_1998}, these surveys have set increasing lower limits on \gls{HZ} dust around nearby stars.
Observing between 8 micron and 13 micron, the Keck Interferometer Nulling experiment placed a 95\% confidence limit of $<60\times$ solar for the median exozodi level around nearby main-sequence stars \citep{mennesson_constraining_2014}.
(
Integrated dust brightness is often quantified by $n_z$, a scalar multiplier applied to Solar System dust distribution, the colloquial ``number of zodis.")
The most sensitive direct search for exozodiacal dust in the \gls{HZ} to date is the Large Binocular Telescope Interferometer  nulling mode \gls{HOSTS}  survey of 10 micron thermal emission \citep{hinz_large_2004,kennedy_exo-zodi_2015,defrere_first-light_2015}. 
\gls{HOSTS} reached a sensitivity below 10$\times$ solar for a number of targets and  \cite{ertel_hosts_2018} inferred a median of 4.5$\times$ the solar zodiacal flux for Sun-like stars. 
This has since been revised to a best-fit median ${\pm}1\sigma$ of 3$^{+6}_{-3}$ zodi, with a 95\% upper limit of 27 zodi  \citep{ertel_hosts_2020}.
The Roman \gls{CGI} sensitivity to exozodiacal light will be considered in detail in Sec. \ref{sec:discussion}.

Providing visible light sensitivity at small separations, the \gls{Roman} Coronagraph (henceforth referred to as the Roman Coronagraph) is expected to detect and spectroscopically resolve reflected light from giant planets \citep{lupu_developing_2016,kasdin_wfirst_2018}, debris disks \citep{schneider_quick_2014,debes_pushing_2019}, and self-luminous exoplanets \citep{lacy_prospects_2020} with predicted sensitivity to companions with flux ratios below 10$^{-8}$ \citep{mennesson_wfirst_2018,bailey_wfirst_2019-1,mennesson_paving_2020}. 
This sensitivity is enabled by an active wavefront control system which corrects for static and dynamic wavefront errors, enabling unprecedented rejection of starlight. 
As a demonstration of exoplanet imaging technology, the high-contrast instrument will also enable scattered light imaging of exozodiacal disks.
This study will quantify the number of habitable zones the \gls{Roman}\footnote{Formerly known as the \gls{WFIRST}.} Coronagraph will be able to search for light scattered by exozodiacal dust, validating and complementing \gls{IR} surveys.
This will be quantified as the sensitivity to dust as a function of $n_z$.
Section \ref{sec:methods} introduces our sample and analysis approach, Section \ref{sec:results} describes the sensitivity of \gls{Roman} to dust in the \gls{HZ} of our sample, and Section \ref{sec:discussion} discusses the implications for future missions as well as areas for further improvement.
\begin{table}[]
    \footnotesize
        \centering
    \begin{tabular}{c|c|c|c|c}
    Symbol & Baseline  & with Model Uncertainty & Threshold Value&  Description\\
    \hline
        $\alpha$ & 2.34 &-&-& Radial Power-law Index\\
        
       $n_z$& 22 m$_\mathrm{V}$arcsec$^{-2}$ &-&-& Surface Brightness at 1 au for $M_\sun$\\
       $d_T$ & 2.363 m &-&-& Telescope Diameter \\
      $\lambda_c$& 575 nm &-&-& Central Wavelength \\
      Bandwidth & 0.10 &-&-& Filter Bandwidth \\
      \acrshort{IWA}  &  0\farcs15&0\farcs165 & 0\farcs28 & \acrlong{IWA}\\
     $t_{\rm occ}^\prime$ &0.12   & .11 & .11 &    Disk Transmission at \acrshort{IWA} \\ 
    $\tau_{\ell}$& 1.4E-11  &2.8E-11 & 1.21E-09 & Fraction Stellar Leakage\\ 
      $\Omega$ &  0.0022 arcsec$^2$ &-&-& Core Area \\ 
    dQE & 0.68  & 0.61 &-&  Quantum Efficiency\\ %
      sread & 0 e$^-$  &-&-&  Read Noise\\ 
      dark & 0.97 [e$^-$/pix/h]  &-&-& Dark Current \\ 
      \acrshort{CIC} & 0.01 e$^-$/pix  &-&-&  Clock-induced charge \\
      $t_{exp}$ & 2  s  &-&-&  Exposure Time \\
      $f_{\Delta I}$ & 0.25  & 0.25 &0.25&  Post-Processing Attenuation\\
    \end{tabular}
    \caption{Key parameters of the EXOSIMS sensitivity model. Baseline instrument contrast and noise values are current best-estimates without performance margins  used by the project for exoplanet yield modeling.
    The project defined model uncertainty values apply padding representing a more conservative case: a 2$\times$ uncertainty in instrument contrast and 10\% uncertainty in \acrshort{IWA} and $r^\prime_{\rm occ}$. 
    Detector performance is projected for 21 months into the mission.
    }
    \label{tab:params}
\end{table}

\section{Methods}\label{sec:methods}
\subsection{Instrument Model}
The \gls{Roman}  Coronagraph narrow-\gls{FOV} mode, nominally using an \gls{HLC},  is expected to be the primary mode for observations of exozodi. 
The \gls{HLC}'s 360 degree dark hole (the coronagraph's high-contrast, high-throughput region) and small ${\sim}3\lambda/D$ \gls{IWA} \citep{trauger_hybrid_2016} make it well suited to imaging circumstellar disks close to a host star.
Another, wide-\gls{FOV} mode, \gls{SPC} with a $6\lambda/D$ \gls{IWA} will provide high-contrast imaging of cool debris disks further from their host star.
Instrument performance from this work is derived from end-to-end diffraction modeling of the coronagraph system, including \gls{STOP} modeling and high-order wavefront control  simulations for fiducial targets \citep{krist_end--end_2014, krist_wfirst_2017,krist_wfirst_2018,zhou_high_2018}. 
These models have been made available by the project to the public\footnote{\url{https://roman.ipac.caltech.edu/}}. 
The model performance predictions were validated against the CGI high-contrast testbed as part of the Roman Coronagraph technology program as described in \cite{zhou_roman_2020,poberezhskiy_roman_2021}.

As described in \cite{krist_wfirst_2018}, to reproduce a realistic coronagraph performance simulation, each \gls{OS} is defined by a set of inputs:  scenes composed of target and reference star, an observatory jitter level, and an assumed orbit and corresponding solar illumination angle.
These inputs are fed into a ``conventional'' observatory \gls{STOP} model which produces wavefront time series from an disturbed optical ray-trace based on positions set by a spacecraft thermal control and structural model.
See \cite{smith_wide-field_2018} for a description of the Roman observatory optical telescope assembly.
The Roman \gls{CGI} is actively controlled, thus the internal \gls{LOWFSC} system after the observatory, described in \citep{shi_low_2016}, is then modeled to remove low-order aberrations such as tip-tilt, defocus, and astigmatism.
The starlight-blocking coronagraph masks \citep{riggs_flight_2021}, high-order diffraction effects, and the contrast improvement provided by the \gls{HOWFSC} \cite{zhou_roman_2020} which measures and removes speckles from the image plane, are modeled in an end-to-end diffraction model \cite{lawrence_optical_1992} using PROPER \cite{krist_proper:_2007}. 
The PROPER model produces a time series of high-contrast images, which show the time evolution of the remaining post-coronagraph speckles in and around the dark hole.

A detector model can optionally be applied to the time series to determine the final contrast for a particular observing time.
OS9 is the ninth and latest public release of Roman post-coronagraph simulated science images, including jitter and an end-to-end \gls{STOP} model of the Roman observatory, coronagraph masks, diffraction,  wavefront control, and detector noise.
OS9 simulates a single simulated target system (47 Uma, $m_{\textrm V}$=5.03) and bright reference star $\zeta$ Pup ($m_{\textrm V}$=2.25). In past \glspl{OS} $\beta$ Uma has been used as a reference (e.g. \cite{ygouf_data_2016}). 

While the integrated STOP model represents the state of the art and the highest fidelity possible for coronagraph performance estimation, it is very time consuming per run, taking as much as a week \citep{krist_wfirst_2018}. 
For many studies such as those involving a large number of targets, this runtime becomes prohibitive.  
The project has developed a comprehensive analytical model of the coronagraph performance, informed by the larger STOP model-derived statistics for any given \gls{OS}. 
In particular, this model has been used to calculate exposure times for known \gls{RV} exoplanets and estimate the sensitivity of those exposure times to instrument parameters \citep{nemati_sensitivity_2017,bailey_wfirst_2019-1}. 
Here we briefly summarize this analytic approach, and provide a few of the key formulae. 


The major categories of error in direct imaging are photometric noise, speckle noise, and calibration errors. Photometric noise includes all of the shot noise sources, including the target, the speckle, and local zodi. It also includes detector noise arising from dark current, clock induced charge, and read noise. Since the Coronagraph uses an \gls{EMCCD}, read noise, which would have otherwise been the dominant noise source, is eliminated, at the cost of some loss of efficiency \citep{nemati_method_2020}.

Calibration errors are those that are incurred when converting the raw counts of signal to flux ratio units, and include such factors as star flux photometric normalization, flat field correction errors, and image photometric correction errors.

The most challenging noise source to model is post-coronagraph stellar leakage, commonly known as speckle noise. This is an important error because it does not tend to go down with integration time and constitutes a noise floor of the photometric measurement. To model this analytically, the Roman Coronagraph project team starts with the the dark hole field contributions from a large number of error sources and with some simplifying assumptions about the level of correlation among the sources.
These are reduced to four summary statistics, relating to the field mean and variance, and their change due to changes in spacecraft pointing, from a target to a reference star, over an ensemble of such observations. 
The sensitivity of these statistics to each type of noise source is computed using the full diffraction model of the coronagraph, and the larger STOP model is used to compute the disturbance statistics.
These statistics are available for each \gls{OS} that has been run using the full model. 
The field statistics that are needed for the speckle noise estimation are then computed using these sensitivities and disturbance estimates from the \gls{OS} runs. 
The analytical model, which also undergirds the Coronagraph error budget and  the official reported performance of the instrument, has been validated against the STOP model and is estimated to produce results that agree with the STOP model within 20\% over most of the dark hole, including near the IWA. 
The most recent output of this analytic model is presented \cite[Figure 8]{mennesson_roman_2021} as $5\sigma$ point source sensitivity  in 100 hours.

The analytical model inputs are arrays of summary statistics as obtained from test data, or models that have been validated and are in formal use by the project system engineering team (for details of these models, see \citealp{krist_numerical_2015,krist_wfirst_2018,douglas_review_2020, poberezhskiy_roman_2021}). 
The key parameters provided by the project and derived from these models are summarized in Table \ref{tab:params}. 
The central wavelength $\lambda_c$ and bandwidth define the near-V-band filter used in narrow-\gls{FOV} mode.
The  fractional stellar leakage, $\tau_\ell$, is the mean intensity fraction of starlight per \gls{resel} of the residual stellar light  (i.e. speckles), in the image plane averaged.  The mean value of $\tau_\ell$ is temporally averaged over the entire observation and calculated for a resel located at the \gls{IWA}.
Assumed detector noise parameters and exposure time are also given in Table \ref{tab:params} for the  photon counting \gls{EMCCD} including 21 months of radiation induced detector degradation \citep{nemati_detector_2014,nemati_sensitivity_2017}.
We adopt the definition of one zodi defined as the surface brightness at 1 AU around a Sun-like star, 22 $m_{\rm V}$arcsec$^{-2}$ \citep{stark_maximizing_2014}. The core throughput is the proportion of light from an off-axis source in the image plane that is transmitted per \gls{resel}, accounting for losses due to the coronagraph and other optics in the system.

\subsubsection{Core-Throughput and post-processing gain}
Two of the main factors distinguishing extended sources and exoplanet observations is calculation of the source throughput per \gls{resel} and the effectiveness of post-processing to remove speckles. 
For an extended, uniform cloud, the ``wings'' of each neighboring \gls{PSF} add to the intensity in the \gls{resel} of interest.
Thus, the photometric correction, for an  extended, source of uniform brightness is significantly elevated versus a lone point source.
$t_{\rm occ}$ is the raw mask transmission of an extended source, such as the local Solar System zodiacal background, measured from end-to-end diffraction models.
Here we use $t_{\rm occ}^\prime=t_{\rm occ}$/2 at the \gls{IWA}, which corresponds to one half of the fractional occulter transmission for an entirely uniform input, to approximate for decreased contribution due to fall off in the disk brightness with radius.
 For space-based observations with relatively stable PSFs, \gls{NMF} is well suited to post-processing of extended sources, as using a nonorthogonal and non-negative basis of \gls{PSF} components preserves morphology and throughput, eliminating the need for forward modeling of post-processing induced attenuation required by more aggressive other algorithms. 
In this work, a post-processing speckle attenuation factor, $f_{\Delta I}$, is used to quantify any post-processing gains and speckle stability, where unity would be raw speckles and zero would be perfect subtraction of the speckles. $f_{\Delta I}=0.25$ is assumed, which is more conservative than the point-source  $f_{\Delta I}\lesssim 0.1$ used for Roman elsewhere (e.g., \citealp{ygouf_data_2016,nemati_sensitivity_2017}), e.g. via \gls{KLIP} and \gls{NMF} \citep{soummer_detection_2012,ren_non-negative_2018,  ygouf_roman_2021}.
As discussed in the appendix, this value was calculated by adding margin to the result of running \gls{NMF} post-processing on \gls{OS} of the star 47 Uma ($m_{\textrm V}$,=5.03) which include both speckle and detector noise processes, made public by the project team. 
This is conservative as post-processing gain is expected on brighter stars where detector noise is less of a factor, as can be seen from the speckle attenuation factor ${\gg}0.1$ found for the noiseless case in the Appendix.


\subsection{Habitable Zone Definition}
The range of habitats in which life could arise is broad.
For detailed discussions, see \cite{kasting_habitable_1993,seager_exoplanet_2013,shields_habitability_2016}.
To capture the variability in stellar irradiance as a function of effective temperature ($T_{\rm eff}$), we adopt the classical \gls{HZ} boundary definitions from Equations 2 and 3 of \cite{kaltenegger_how_2017}, which relates the effective stellar flux $S_{\rm eff}$ to $T_{\rm eff}$ for A- through M-type stars.
This relationship is captured via a third order polynomial fit to one dimensional atmospheric models which include greenhouse effects and geochemical cycles that regulate atmospheric CO$_2$. 
To capture a wide range of possible habitable zones, this analysis uses the polynomial constants \cite[Table 2]{kaltenegger_how_2017} for the outer edge derived for an early Mars and the inner edge for a recent Venus conditions  derived from \cite{kopparapu_habitable_2013} and \cite{ramirez_habitable_2016}. 
As a simplification, to estimate an Earthlike position within the \gls{HZ} while accounting for varying stellar luminosity, the \gls{EEID} is the distance from the star where the incident energy density is that of the Earth received from the Sun. 

\subsection{Targets}
As a representative sample of targets, we choose the HabEx mission target list of nearby main sequence stars \citep[Appendix D]{gaudi_habitable_2020}. 
That list of highest priority targets was created for a HabEx design reference mission of 5 years, assuming no prior knowledge of exo-Earth candidates, splitting the observing time between exo-Earth searches and orbit determination with HabEx coronagraph (around 0.5 micron), followed by spectral characterization of  with HabEx starshade (0.30–1.00 micron).
Target stars -- as well as visiting epochs and observations durations -- are determined by applying the altruistic yield optimization \citep{stark_lower_2015} algorithm to maximize the number of exo-Earths characterized by the mission over 5 years. 
This list is representative of the target sample for a direct imaging mission such as HabEx or the \gls{LUVOIR} concept \citep{the_luvoir_team_luvoir_2019}. 

The EXOSIMS architecture
\citep{savransky_wfirst-afta_2016,savransky_exosims:_2017} was used to calculate each target's surface brightness at the coronagraph \gls{IWA}  using the formalism  described by \cite{nemati_sensitivity_2017,nemati_method_2020}. After exclusion of binaries, 149 targets were identified and stellar properties were retrieved from Simbad to calculate the \gls{HZ} location and relevant photon rates. 
Assuming a Solar-like distribution of dust, the exozodiacal brightness is approximated as a function of the radial distance from the star.
Abbreviated as $f_{\rm EZ}$, this takes the form of a flux, $\phi$, per solid angle, $\Omega$:

\begin{equation}
f_{\rm EZ}(d) =\frac{\partial \phi_{\rm EZ}}{\partial \Omega}(d)=
 \left[\frac{\partial \phi}{\partial \Omega}(_{\rm 1AU})\right]\left[10^{M_\star-M_\sun}\left(\frac{1}{d}\right)^{\alpha}\right]\label{eq:fez}.
\end{equation}
Where the normalization term, 
\begin{equation}
  \left[\frac{\partial \phi}{\partial \Omega}(_{\rm 1AU})\right]\equiv m_z  =22
  \end{equation}
  and has units of ${\textrm {mag}}$\ arcsecs$-^2$.
$M_\star$ and $M_\sun$ are bolometric magnitudes to correct for differences in stellar luminosity \citep{stark_maximizing_2014}.
$d$ is the radial separation distance from the star in AU, and $\alpha$ is the power law index of surface brightness as a function of radius. 
We adopt $\alpha=$ 2.34, derived from \cite{kelsall_cobe_1998} and used by \gls{HOSTS} \citep{kennedy_exo-zodi_2015}, to account for both the decrease in incident flux and the decrease in density. 


The exozodi count rate per \gls{resel} is given by: 

\begin{equation}
r_{z} = n_z \Omega t_{\rm occ}^\prime f_{\rm EZ}(d).
\end{equation}{}\label{eq:r_z1}

Here $t_{\rm occ}^\prime$ is the per \gls{resel} core throughput, and $\Omega$ is the solid angle of the core of a post-coronagraph \gls{resel}.
By default EXOSIMS calculates an inclination dependent brightness correction $f(\beta)$, see \cite{savransky_analyzing_2009}; for this work, we conservatively set this value to unity, which corresponds to a face-on-disk and the lowest peak-brightness scattering geometry. 
 



\subsection{Sensitivity}
A challenge of high-contrast imaging is accurately accounting for speckle noise and post-processing routines.
Speckle subtraction for exoplanet detection often relies on multiple roll angles on a single star (\gls{ADI} \citep{lafreniere_new_2007}). 
This generally leads to self-subtraction of symmetric extended structures. Thus, subtraction of reference PSF  libraries, \gls{RDI} \citep{lafreniere_hst_2009}, from dustless stars is generally required to recover detect disks (e.g. \cite{schneider_probing_2014,choquet_first_2017}). 
Adopting the same assumptions with respect to speckle behavior as Roman exoplanet sensitivity work,  expression for photometric \gls{SNR} for \gls{RDI}, $R$, in a \gls{resel} in a coronagraph from \cite[Eq. 74]{nemati_method_2020}:

\begin{equation}
R=\frac{r_\phi \ t}{\sqrt{r_{\rm n}t+f_{\Delta I}^2 \ r_{\rm sp}^2t^2}}\label{eq:SNR}
\end{equation}
where we have replaced the planet flux $r_p$ with $r_\phi$ to represent the exozodiacal signal rate, in photos per second per \gls{resel}, $r_{\rm n} $ is the noise rate, including photon shot noise, detector noise, and background sources, and $r_{\rm sp}$ is the residual speckle rate. 
 We multiply this by $f_{\Delta I}$ and solve Equation~\eqref{eq:SNR} for exposure time as a function of \gls{SNR}, $R$, giving
\begin{equation}
    t_{\rm S}= \frac{R^2\ r_{\rm n}}{r_\phi^2-R^2\ f_{\Delta I}^2 \  r_{\rm sp}^2 } .
\end{equation}\label{eq:t_snr}
For long exposure times,  $t \gg r_{\rm n}/(f_{\Delta I}r_{sp})^2\sim\infty$, the detectable signal depends only the residual speckle rate:   $r_\phi=(R\ f_{\Delta I}\ r_{sp})$. 
Here we assume all gains due to reference subtraction and post-processing are captured in $f_{\Delta I}$ and additional speckle information provides negligible gain, simplifying the analysis but potentially under-estimating the fundamental limit of long observations relative to the speckle lifetime (see discussions in e.g., \citealp{males_orbital_2015, young_image_2013}).


Assuming a solar-like zodiacal dust cloud multiplied by a scalar factor:
\begin{equation}
    n_{z}= \frac{ r_{sp}\ f_{\Delta I} \ R}{f_{\rm EZ}(d) \Omega t_{\rm occ}^\prime f_{\rm EZ}(d)},\label{eq:n_critical}
\end{equation}
where $R$ is the desired \gls{SNR} and $F_{\rm EZ}$ is the flux per \gls{resel} of a solar-system zodi. 
This formulation is conservative because we would expect to detect most debris disks in multiple \glspl{resel}.
Since we focus on the brightest point in the disk, right at the \gls{IWA}, we assume a detection threshold of $R=5$. 
For a flat speckle brightness as a function of separation, other image elements will have scattered light above the speckle floor but below $5\sigma$, allowing increased certainty of detection and decreasing the likelihood of coronagraph leakage confounding the detection. 
Since speckles are also expected to decrease with radius from the star, albeit slower than disk brightness, this is also a conservative assumption.


\section{Results}\label{sec:results}

\begin{figure}
    \centering
    \includegraphics[width=.6\textwidth]{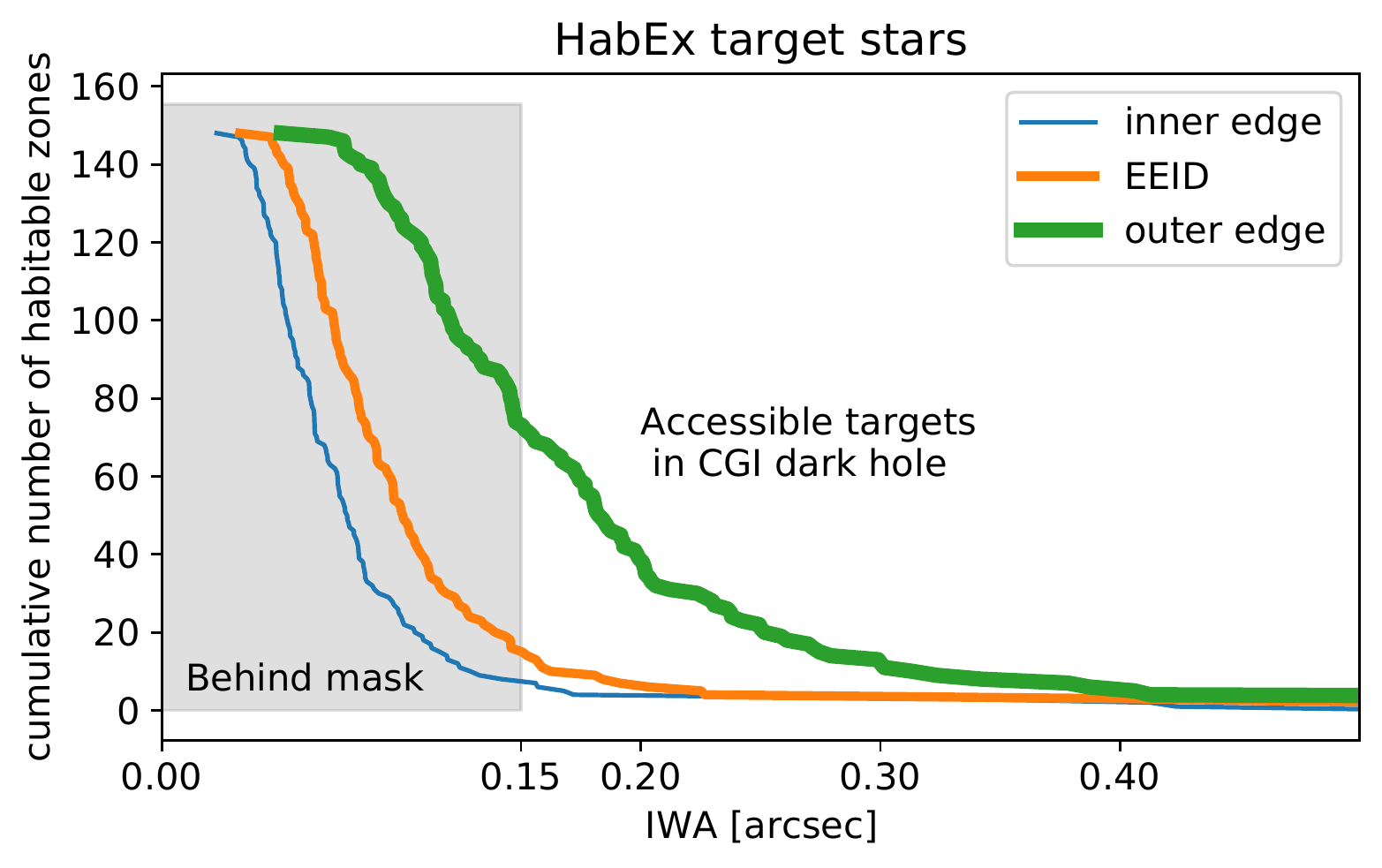}
    \caption{Cumulative  number of HabEx target stars as a function of coronagraph \acrshort{IWA}.     The unobserved region obscured by the Roman Coronagraph \gls{HLC} mask \gls{IWA} is shaded. Systems in the unshaded reagion Roman can either access the habitable zone outer edge (top, thick green curve), earth-equivalent isolation distance (middle, orange curve) or the HZ inner edge (bottom, thin blue curve). 
    The thick top curve corresponds to the young Mars edge while the bottom thin curve corresponds to the recent Venus inner edge.
For this sample, 74 stars have at least a part of their \gls{HZ} accessible, including 16 observable down to their \acrshort{EEID}. }        \label{fig:n_hz}
\end{figure}

\begin{figure}
\includegraphics[width=0.7\textwidth]{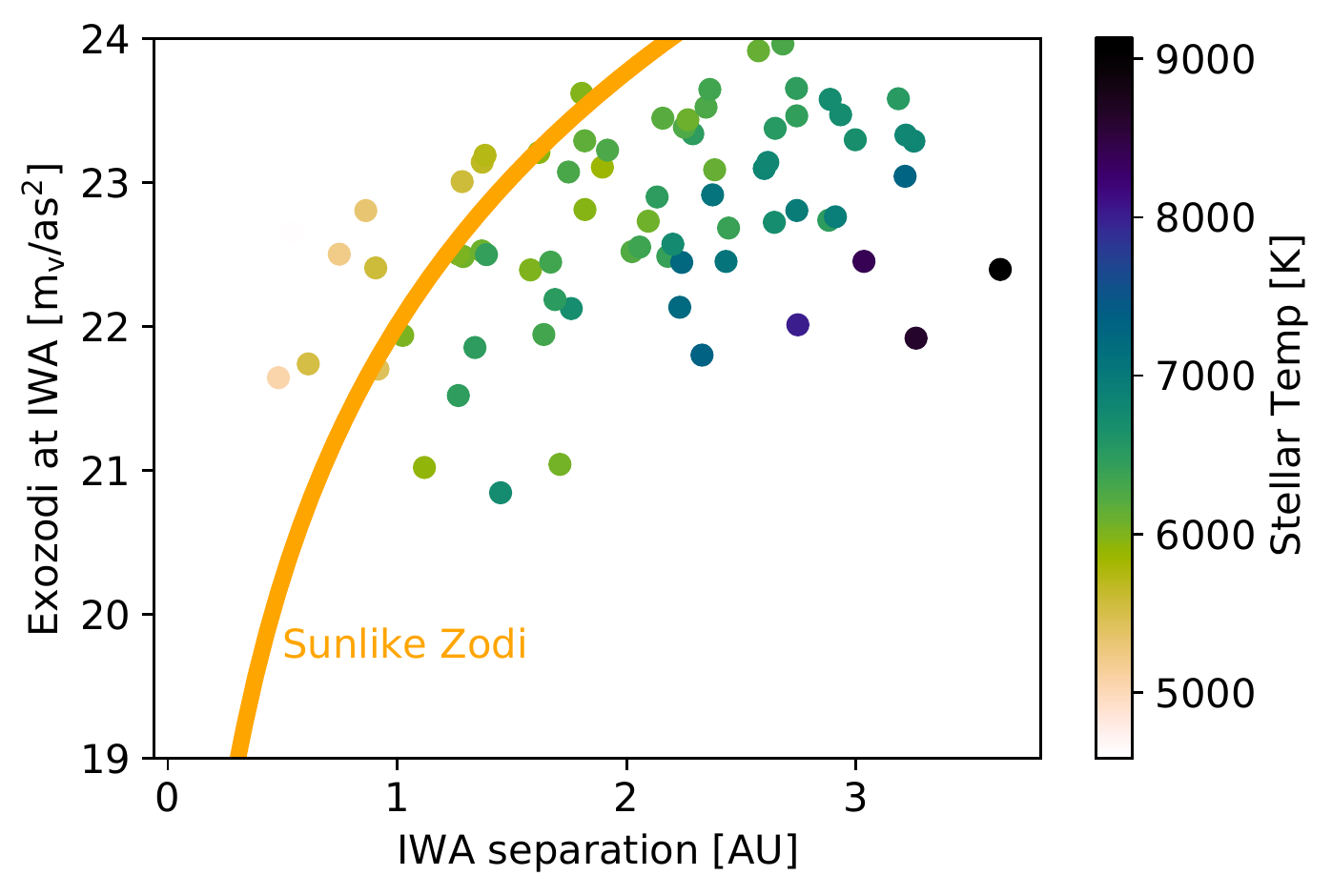}
    \caption{Target population surface brightness versus separation. The solid line shows the assumed surface brightness of a Solar-system-like exozodiacal as function of distance from the host star. 
          Individual points show visible \gls{HZ}'s surface brightness at the \acrshort{IWA} and are colored by stellar temperature. 
          Selection biases make the overall sample  hotter than a Sunlike star.
          }\label{fig:IWA}
\end{figure}


\begin{figure}
\gridline{\fig{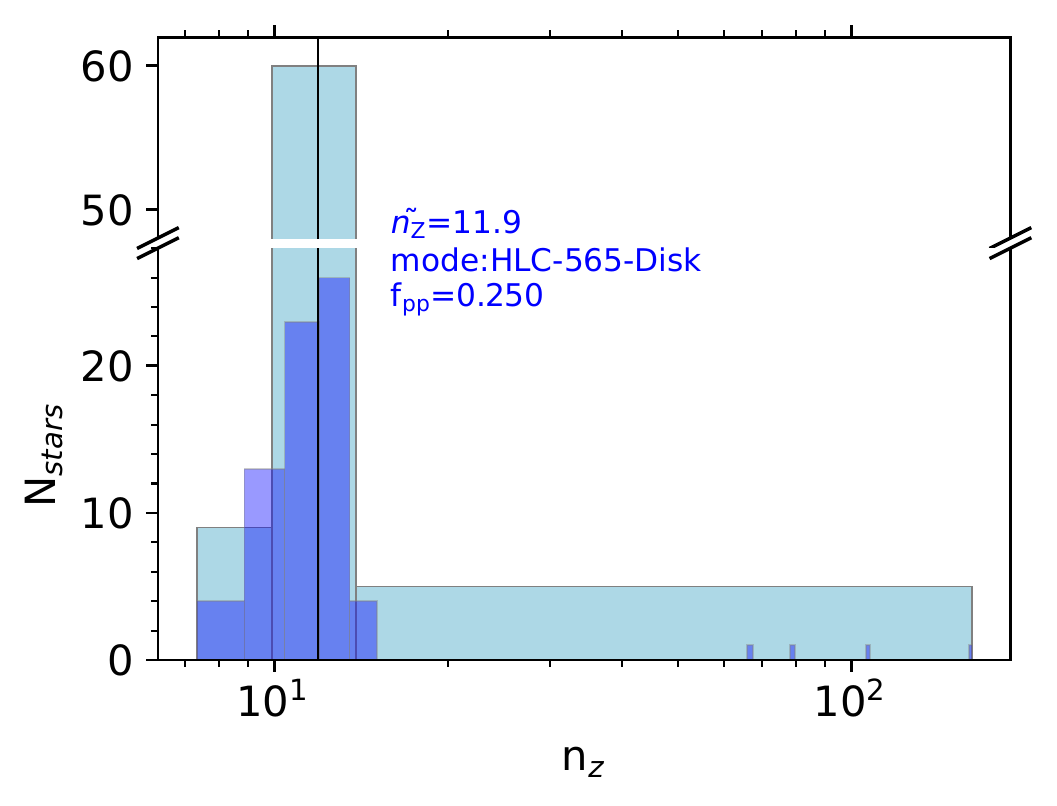}{0.4\textwidth}{(a)}\fig{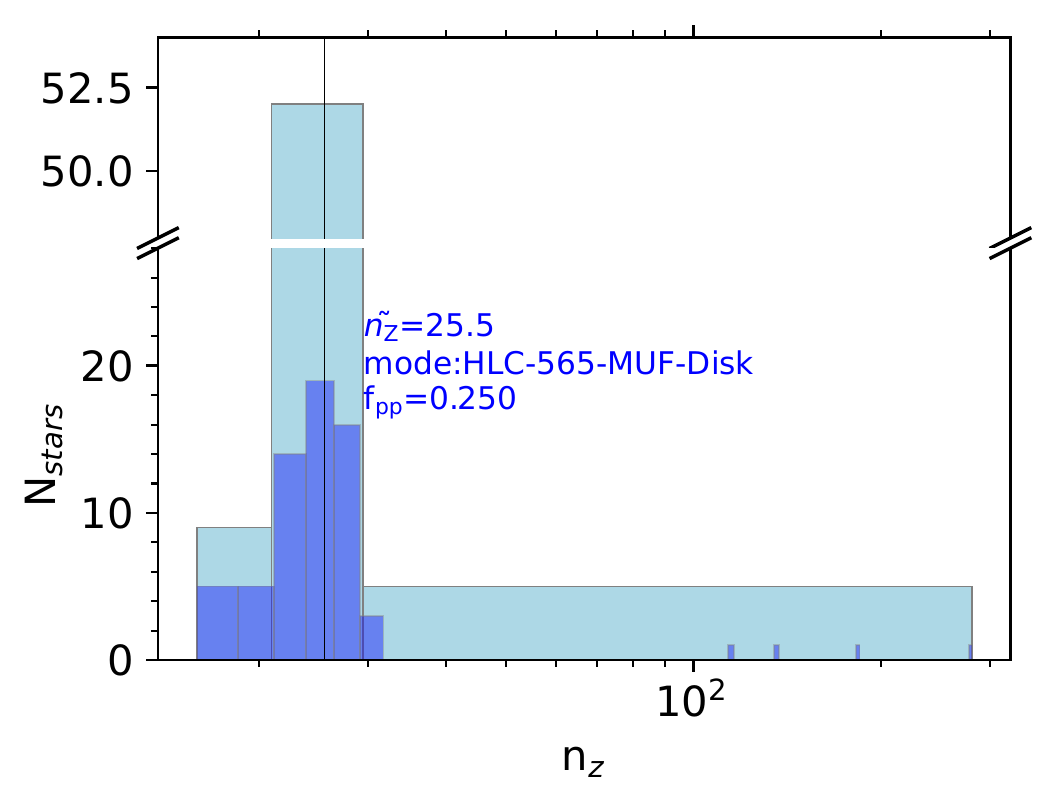}{0.4\textwidth}{(b)}}
    \caption{Distribution of $5\sigma$ exozodi sensitivities for  stars in the sample in the speckle limited, long exposure time, case. Narrow, dark bins correspond to evenly spaced bins;  wide, lightly shaded bin sizes are set dynamically using the Bayesian Blocks method to show only significant differences in the distribution \citep{scargle_studies_2013}.
(a): current best-estimate median exozodiacal sensitivity where $\tilde{n_Z}$ is the calculated median number of zodi;
(b): distribution of critical exozodi sensitivities for stars in the sample, 
for the ``model uncertainty factor'' (MUF) padded level of instrument performance, where the \gls{IWA} increases to 0\farcs165 and the flux ratio sensitivity (contrast) degrades by a factor of two, the latter dominating the decrease in median sensitivity.
}\label{fig:sensitivity}
\end{figure}

%
%

Geometry sets limits on the sample of \gls{HZ} that \gls{Roman} can observe at high-contrasts. 
Considering the exozodi around each star at the inner and outer edges of the \gls{HZ} brackets the region of interest.
While the surface brightness at the inner edge is greater, the outer edge is more likely to fall outside the \gls{IWA}, leading to significantly more detections.  
This is seen in Figure~\ref{fig:n_hz}, where the cumulative number of  accessible habitable zones is shown as a function of coronagraphic inner working angle.
The number of \gls{HZ} outer edges (thick green line) inside the Coronagraph dark hole is 74 and eight inner edges (thin blue line) are visible, assuming the nominal 0\farcs15 \gls{IWA}.
Sixteen systems have an  \gls{EEID} inside the coronagraph dark hole (medium orange line).
As stars get more distant, the coronagraph IWA translates to a larger physical separation that only remains within the HZ of the hottest  and most luminous stars in the sample.
This gradient in temperature is visible in Figure~\ref{fig:IWA}, which shows the surface brightness, the \gls{IWA} separation in AU for each star with a visible \gls{HZ}.
The orange curve shows the surface brightness of the archetypal Solar zodiacal disk around a Sunlike star.
Since there is a selection bias towards more luminous stars, the predicted exozodiacal surface brightness of the sample of geometrically observable \gls{HZ} tends to be brighter than the solar zodiacal light at a given separation (see equation 1 with $M_\star > M_\sun$).


To quantify the exozodi sensitivity for each star in the sample, the critical  sensitivity (Equation~\eqref{eq:n_critical}) is calculated at the \gls{IWA} for each star's V-band magnitude and distance.
Figure~\ref{fig:sensitivity} shows a histogram of the number of stars observed as function of $n_z$, in the long exposure time limit.
Because of the fixed \gls{IWA} of the Roman coronagraph,  the average star amenable to exozodi observations within the HZ is hotter than the Sun.
And its exozodi surface brightness at a given physical separation is larger than expected around a Sun-like star (as shown in Figure~\ref{fig:IWA}).
Since the irradiance received by the \gls{HZ} is a constant by definition, these hotter stars have more widely separated \gls{HZ}. 
increasing with stellar distance and since both density and irradiance are dropping the flux decreases  $\propto d^{-\alpha}$, as shown in Equation~\eqref{eq:fez}.
The long tail in Figure~\ref{fig:sensitivity}  above 60 zodi is made up of the four stars in the sample with the brightest apparent magnitude:  Procyon, Altair, Fomalhaut, and Denebola. 
These stars have correspondingly higher speckle brightness relative to predicted exozodiacal dust surface brightness at the \gls{IWA}, though they may be detectable at higher sensitivities if observed at larger separations.
This is reflected by the spread of values in Figure~\ref{fig:sensitivity} around the median value of {$\sim$12} zodi, which does not include uncertainty in instrumental sensitivities.

\begin{figure}[htbp]
\begin{center}
\includegraphics[width=.65\textwidth]{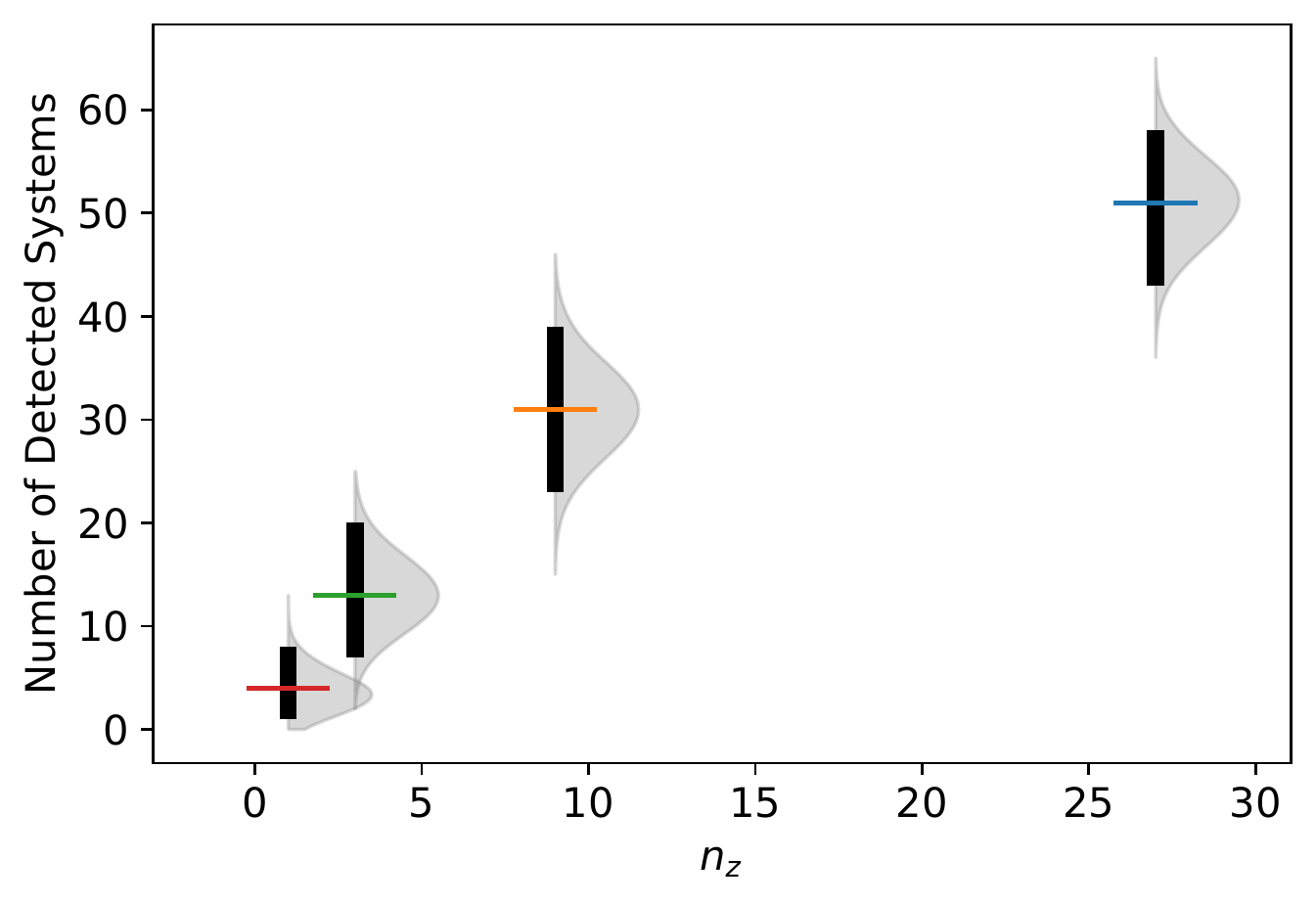}
\caption{Expected number of systems with \gls{HZ} exozodiacal dust detected by the Roman Coronagraph as a function of increasing median exozodi level ( $n_z$ in solar units). 
    The y-axis represents  $5\sigma$ exozodi detection for stars in the sample in speckle-limited case. 
    The distributions of $5\sigma$ detections for the \gls{CBE} sensitivity derived from randomly drawn log-normal distributions of increasing median-$n_z$ and $\varsigma$=1.5.
    Vertical bars represent the 95\% percent confidence intervals, shaded regions show the underlying probability distributions, and horizontal bars indicate the median.
 For the \gls{HOSTS} predicted median of 3 zodi, there is a $>95\%$ confidence that multiple systems will be detectable.
}
\label{fig:crit_dist}
\end{center}
\end{figure}

To estimate the number of systems where exozodi may be detected at any level, we chose four representative values of $n_z$, unity (i.e. Solar) and the nominal median exozodi level, plus the $1\sigma$, and 95\% confidence upper limits derived from the  \gls{HOSTS} survey \citep{ertel_hosts_2020}.
(The $1\sigma$ HOSTS lower limit is zero). 
For each system on the target list, we drew increasing numbers of possible realizations of the log-normal distribution (with a standard deviation of $\varsigma=1.5$, the value found by \cite{ertel_hosts_2018-1} for dust-less stars).
The results stabilized after approximately 10,000 independent and identically distributed  draws per case. 
The number of outer \gls{HZ} exozodi which might be detectable for increasing values of $n_z$ are shown as half-``violin plots'' in Figure~\ref{fig:crit_dist}.
$n_z$ shown correspond to one zodi, the three zodi median  from \cite{ertel_hosts_2020}, as well as the 1$\sigma$ (nine zodi) and 95\% confidence levels (27 zodi).
Shaded regions' widths indicate the underlying distribution, horizontal lines indicate the median and thin vertical lines represent 95\% confidence intervals.
Table \ref{tab:ndet} provides a summary of the same median exozodi levels with median and 95\% confidence intervals.
This assumed distribution is conservative, as will be discussed below.

\begin{figure}
    \centering
    \includegraphics[width=.6\textwidth]{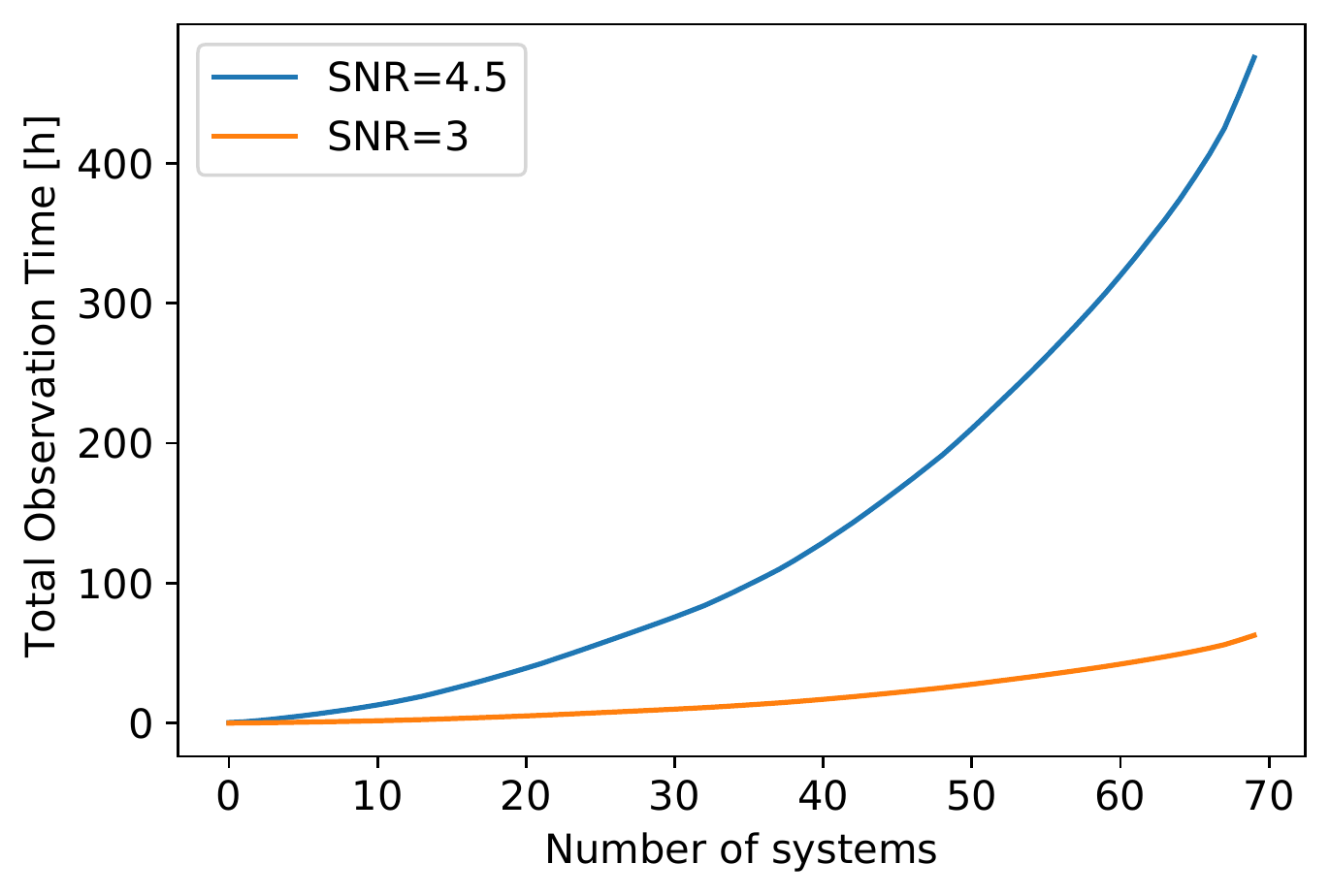}
    \caption{
Number of observable systems versus cumulative mission exposure time, neglecting acquisition time or visibility, for the \gls{CBE} sensitivity and detector properties. 
Each system is observed until the critical exozodi sensitivity is reached at $S=4.5$, since an SNR of five, by definition, is only  reached at the critical dust level in an infinite exposure time. 
A 3$\sigma$ detection sensitivity exposure time is shown in the lower curve. A  3$\sigma$ detection in 3 \glspl{resel}, a likely scenario for most geometries, provides a $5\sigma$ detection of the presence of a disks around most stars in a significantly shorter survey. }\label{fig:exptime}
\end{figure}

Thus far, we have counted detections in the long exposure time regime. 
To estimate the minimum time for a survey,
Equation~\eqref{eq:t_snr} gives the exposure time as function of source flux, noise rate, and speckle rate. 
Applying this to our sample, Figure~\ref{fig:exptime} shows the cumulative exposure time for the sample, given by Eq. \ref{eq:t_snr}, is hundreds of hours. 
The entire sample is likely observable to high SNR-per-resel in a few weeks time, albeit spread throughout the sidereal year and with additional observing overheads.
Importantly, the second curve shown on Figure~\ref{fig:exptime} demonstrates a much shorter exposure time is necessary to reach \gls{SNR}=3 per \gls{resel}, which will allow significant detection of extended disk structures quickly in multiple \glspl{resel}.

\begin{table}[]
    \centering
      \begin{tabular}{c|c|c|c}
        $n_{z}$ & median & 95\% Confidence \\
        \hline
    \input{conf_tab_values2_34alphaHLC-565-Disk.tex}
    \end{tabular}
    \caption{The number of systems that can be detected at a 5$\sigma$/resel cut-off, assuming infinite exposure time, a log-normal distribution, $\varsigma$=1.5 and median exozodi levels $n_z$, neglecting uncertainties in the instrument performance model.}
    \label{tab:ndet}
\end{table}

 \begin{figure}
    \centering
    \includegraphics[width=.6\textwidth]{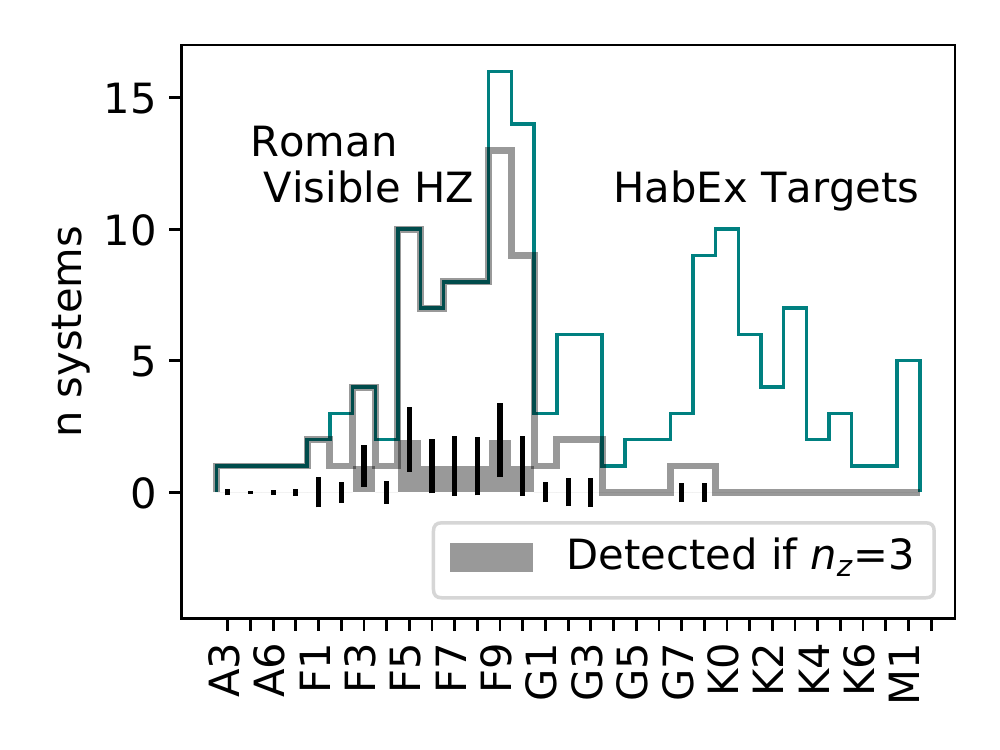}
    \caption{Histograms showing number of stars versus spectral type.
    Shaded region shows the median number of detected systems and bars indicated $1\sigma$ uncertainty calculated from the log-norm draws for the 3 zodi median case.
    The visible Roman HZ (medium width line) and the  HabEx targets (thin line) peak at spectral type F9 but the Roman visible HZ go to zero at G8 and later-type.
    95\% of the HabEx targets (thin line) are FGK stars and all of the Roman predicted detections are FG stars.
  }\label{fig:spectraltypes}
    \end{figure}
\section{Discussion}\label{sec:discussion}

This analysis has shown the \gls{Roman} Coronagraph will place new limits on scattered light brightness from exozodiacal dust in the \gls{HZ} of nearby stars. 
Such a program would provide valuable insight into the scattered light background faced by future missions to image and spectrally characterize Earthlike planets.
Simultaneously, such observations will increase our understanding of exozodiacal dynamical  processes and constrain the relationship between \gls{IR} observations of disk thermal emission and the light scattering albedo. 
As mentioned in \cite{ertel_hosts_2020}, foreknowledge of which systems have excess exozodiacal light will allow better optimization of future direct imaging searches for Earthlike planets.
Thus, both detections and upper limits on the brightness of scattered light at the \gls{CBE} sensitivity decrease the uncertainty in  both  the median brightness and the brightness around specific stars.
This has the potential to optimize the exoplanet observing strategy of future missions by allowing selective targeting of less dusty systems and, perhaps more importantly, informing the aperture required to detect Earthlike exoplanets and record their spectra at a useful \gls{SNR}.

The results presented here retain the log-norm distribution as a conservative estimate of the Coronagraph exozodi detection rate, shown Figure~\ref{fig:crit_dist}.
The 3 zodi distribution represents the median of the underlying exozodiacal brightness distribution reported by \cite{ertel_hosts_2020}; however, the authors of that study found a log-norm distribution did not well fit the exozodiacal brightness function observed by \gls{HOSTS} and instead chose to characterize the distribution with a free-form fit. 
 \gls{HOSTS} results show a subset of systems are dustier than the 3 zodi log-norm would predict, and hence may be easier to detect with the \gls{CBE} performance of Coronagraph. 
\gls{HOSTS} reported median $1\sigma$ \textit{sensitivity} is 48 zodis for Sun-like stars and 23 zodis for early-type stars (A-F5).
As seen in Fig. \ref{fig:spectraltypes} the habitable zones visible to the Coronagaph straddle these two populations, largely made up of spectral types F5-G1, with the majority of the the exozodi detected in the 3 zodi log-normal case around stars  F6 or redder.
While the two populations studied are not identical, the \gls{CBE} Roman median sensitivity (Figure~\ref{fig:sensitivity}(a)) is expected to be more than at least a factor of $\sim2\times$ HOSTS and order of magnitude increase compared to \gls{HST} \cite{debes_pushing_2019-1}. 

%

As a technology demo, the minimum performance of the Roman Coronagraph is set by threshold requirements \citep{douglas_wfirst_2018} that are significantly more relaxed than the current-best-estimate sensitivity presented here. 
At threshold performance level, the \gls{IWA} moves out to 0\farcs28,    $t_{\rm occ}$ decreases to 0.05 and $\tau_{\ell}$ increases to 1.21E-9.
 Re-running the analysis above, while holding $f_{\Delta I}$ and $t_{\rm occ}^\prime$ constant, the  $5\sigma$ exozodi sensitivity limit for the threshold mission sensitivity is of order $10,000$ zodi for 14 observable systems.
This sensitivity to exozodiacal dust is implausibly low, since the increased photon rate would be expected to dramatically increase recovery of speckles and decrease $f_{\Delta I}$ from the assumed value. 
In this case,  $f_{\Delta I}$ would approach the noiseless values of 0.01 to 0.05 (see appendix) but lacking detailed simulations of the instrument performance in the threshold performance regime we do not assert a predicted threshold performance sensitivity. 
Given the conservatism of the analytic estimates, on-orbit performance is expected to meet or exceed the \gls{CBE} results presented, with the ``model uncertainty" sensitivity of 25 zodi providing a likely lower limit.
In particular, as shown in the appendix, when adding ``model uncertainty" leakage  the speckles are better attenuated, suggesting increased post-processing gain is a function of speckle brightness and brighter stars will also have brighter speckles for equal coronagraph performance. 
The median star in the Roman visible HZ sample is nearly 2$\times$ brighter than 47 Uma, suggesting additional post-processing gains may be achievable.


Future work is required to quantify the dependence of Coronagraph sensitivity to a range of higher order effects, including:  
morphological variation such as narrow rings or clumps \citep{defrere_nulling_2010},
partial transmission of light inside the \gls{IWA}  \citep{milani_faster_2020}, 
higher-order calibration and detector noise effects \citep{nemati_detector_2014}, 
degeneracy between coronagraph leakage and disk morphology (particularly low-order aberrations), 
 variations in composition and albedo \citep{debes_cold_2019},
 sensitivity improvements from to matched filtering \citep{defrere_direct_2012},
and dependence of post-processing gain on speckle lifetimes.
Another phenomenon that the Coronagraph will be sensitive to is pseudo-zodi, where forward scattering from inclined asteroid belt analogs increases the background flux that appears close to the star  \citep{stark_pseudo-zodi_2015}.

Mid-infrared observations, such as with \gls{LBTI} and including non-detections suggesting the dust is cool, will help disambiguate these cases.
Currently, the \gls{Roman} mission lifetime is fixed and the Coronagraph technology demonstration mission is currently planned for 3 months duration \citep{bailey_wfirst_2019-1}.
Detailed mission modeling will be required to establish the number of systems which can be observed in hypothetical Coronagraph observations following the technology demonstration phase.

Code to reproduce the figures presented in this study is available on github.com\footnote{\url{https://github.com/douglase/exozodi_exosims_sensitivity}} and archived on Zenodo \citep{douglas_exozodi_exosims_sensitivity_2020}.


\acknowledgements
The authors acknowledge valuable inputs from the JPL and IPAC Roman Coronagraph teams. Thanks to Vanessa Bailey and Steve Ertel for helpful feedback.
 Portions of this work were supported by the WFIRST Science Investigation team prime award \#NNG16PJ24C.
 Portions of this work were supported by the Arizona Board of Regents Technology Research Initiative Fund (TRIF).
J.A.: this work was supported by a NASA Space Technology Graduate Research Opportunity.
 This research has made use of the SIMBAD database and the VizieR catalogue access tool, both operated at CDS, Strasbourg, France.  
 This research has made use of the NASA Exoplanet Archive, which is operated by the California Institute of Technology, under contract with the National Aeronautics and Space Administration under the Exoplanet Exploration Program. 
\facilities{SIMBAD,VizieR, Exoplanet Archive} 
\software{This research made use of community-developed core Python packages, including: Astroquery \citep{adam_ginsburg_astropy/astroquery:_2018}, Astropy \citep{the_astropy_collaboration_astropy:_2013}, Matplotlib \citep{hunter_matplotlib:_2007}, SciPy \citep{jones_scipy:_2001}, Jupyter and the IPython Interactive Computing architecture \citep{perez_ipython:_2007,kluyver_jupyter_2016}. 
Specific to exoplanet imaging, this research made use of the \texttt{EXOSIMS} exoplanet mission simulation package \citep{savransky_exosims:_2017};
for photon-counting, \texttt{EMCCD Detect} \footnote{\url{https://github.com/wfirst-cgi/emccd_detect}}, based on \cite{nemati_photoncounting_2020};
 and for post-processing, the dimensionality reduction code for images using vectorized Nonnegative Matrix Factorization (NMF) in Python \citep{zhu_nonnegative_2016,ren_non-negative_2018,ren_nmf_imaging_2020}. }

\begin{figure}[h]
    \centering
    \includegraphics[width=0.5\textwidth]{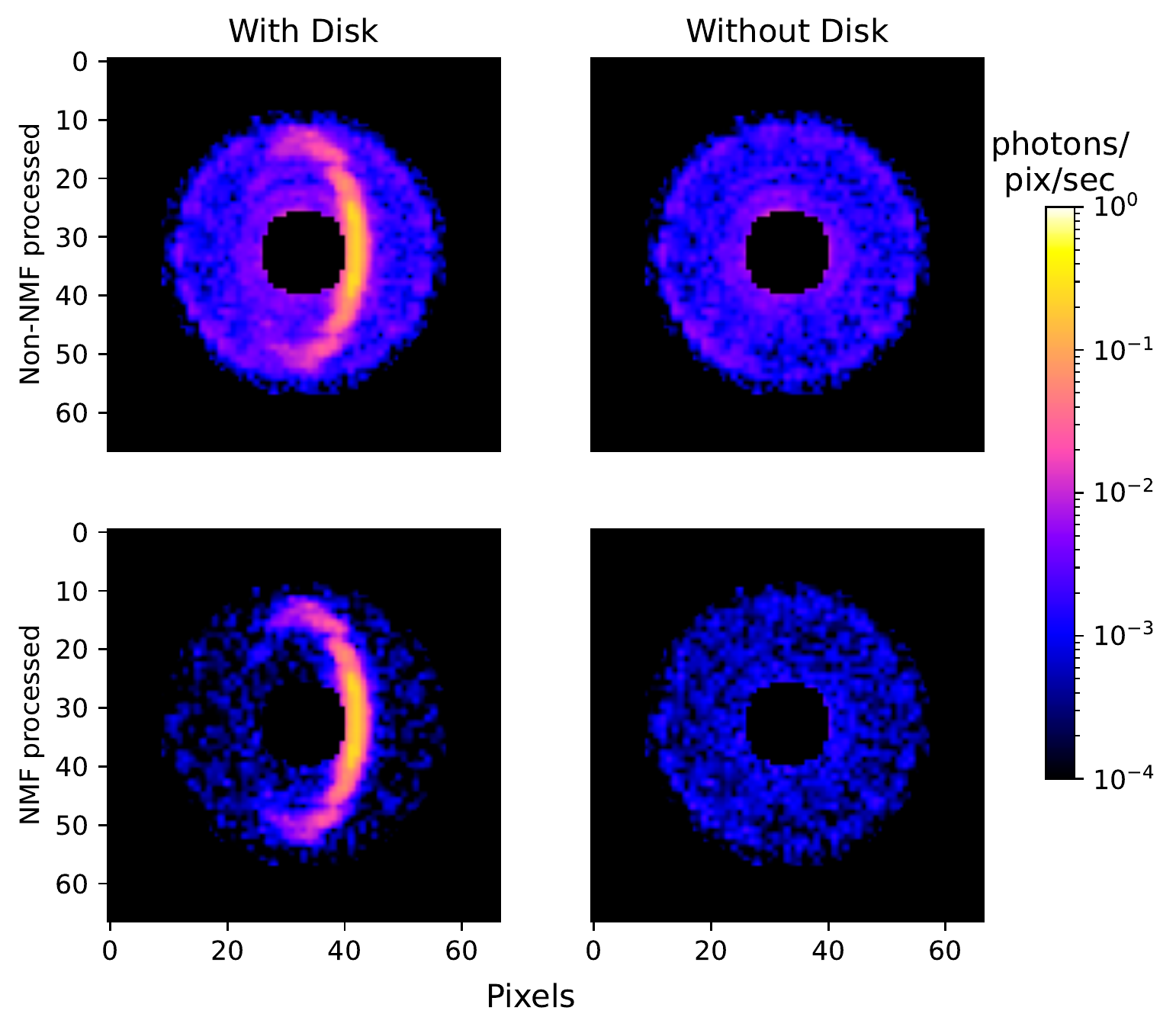}
    \caption{Comparison of the unprocessed (top) and processed (bottom) OS9 frames (without MUFs) with (left) and without (right) an inclined disk injected to demonstrate NMF's ability to subtract speckles from the stellar PSF. 
    Since self-subtraction is generally negligible for \gls{NMF} post-processing, the speckle attenuation factor is given by taking the mean value of the ratio of the top right and bottom right images.}
    \label{fig:NMFquad}
\end{figure}

\newpage
\bibliography{exoplanets.bib,inprep.bib}
\appendix{Simulated Disk Post-Processing:}
The speckle attenuation factor was derived from post-processing of the Roman-Coronagraph HLC Observing Scenario 9 data (OS9\footnote{\url{https://roman.ipac.caltech.edu/sims/Coronagraph_public_images.html}}).
This analysis used the \gls{HLC} model with and without MUFs (margin of uncertainty factors).
The processing procedure begins with photon counting both reference and target frames using the procedure outlined in \cite{nemati_photoncounting_2020}. To subtract the PSF we apply the NMF method of \cite{ren_non-negative_2018} to construct PSF components from the OS9 reference frames. These were then subtracted from the target frames to arrive at a final processed frame. The ratio of the NMF-subtracted frames to the  unprocessed frames near the IWA gives our speckle attenuation factor. To simulate cases with an edge-on disk Nemati's \texttt{EMCCD Detect} code was used to simulate the response of an debris disk model (described in \cite{mennesson_wfirst_2018} and publicly available\footnote{\url{https://roman.ipac.caltech.edu/sims/Circumstellar_Disk_Sims.html}}) on Roman's \gls{EMCCD}. 
These were then added to each OS9 frame before the aforementioned photon counting procedure. 
The signal-to-noise ratio was computed by averaging the median per-pixel frame of the cases without disks for the processed and unprocessed cases separately These are shown in Figure~\ref{fig:NMFquad}.
As a test of the NMF algorithm, an example inclined disk was inserted into the raw frames and is well recovered (bottom left).

\begin{figure}
\gridline{\fig{speckle_atten_noiseless.pdf}{0.4\textwidth}{(a)}
\fig{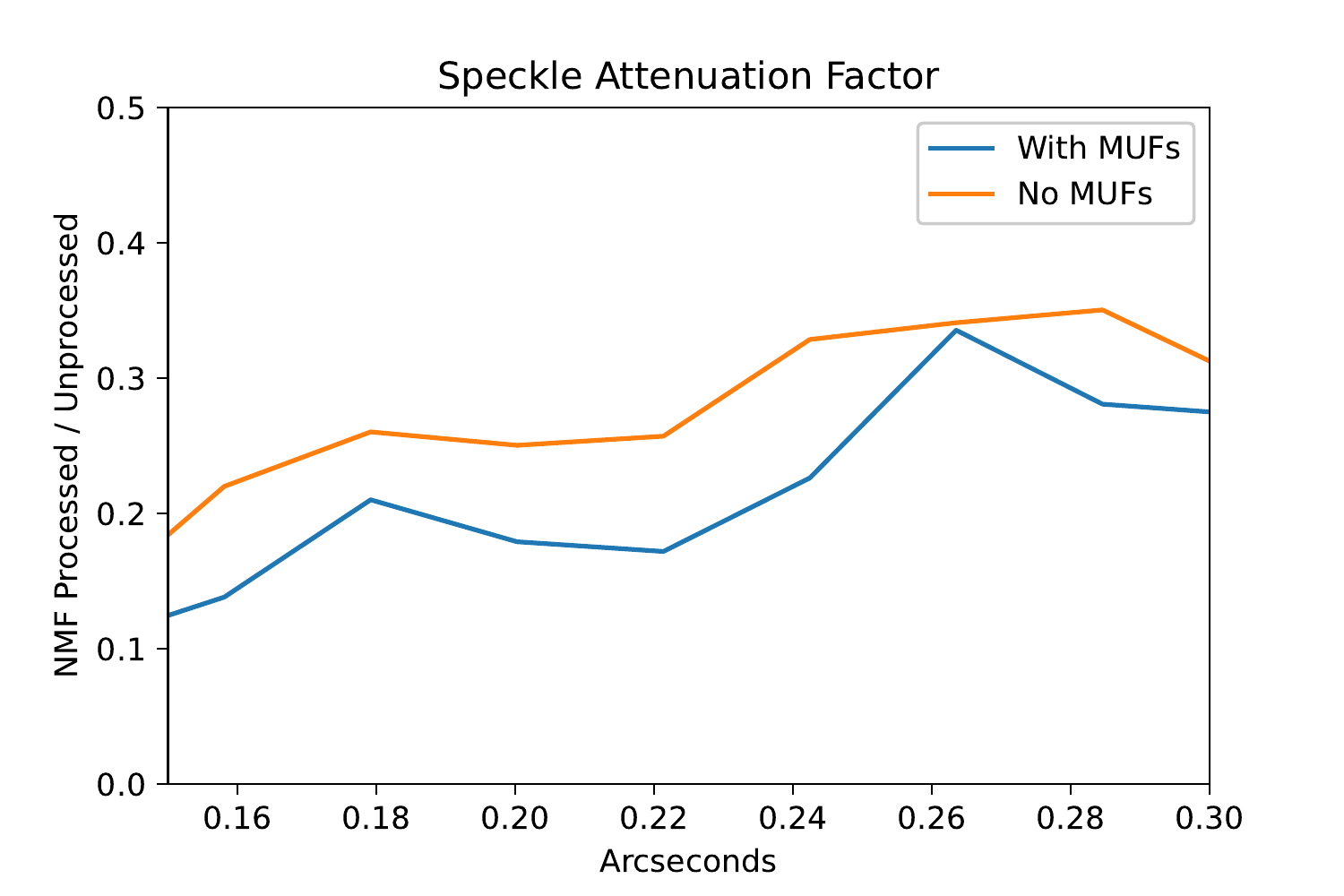}{0.4\textwidth}{(b)}}
   \caption{ \gls{NMF} post-processing  speckle attenuation factor  $f_{\Delta I}$, calculated as the ratio of the processed to unprocessed speckles  as a function of radius for OS9 simulated observations of the 47 Uma system.
   The post-processing gain depends both on the temporal behavior of speckles at certain spatial frequencies and how well the speckles are resolved versus the noise floor. For the  noiseless (a) only spatial frequency and sampling effects apply. 
   When including detector and photon noise (b: coronagraph leakage is dominated by low-spatial frequencies, and speckles close to the \gls{IWA} are better subtracted. 
   Further from the star, detector noise begins to dominate and the attenuation factor flattens out. 
   This  simulation informed our choice of  $f_{\Delta I}$  to be 0.25, a conservative  approximation  of the noisy simulation at 0\farcs15- 0\farcs20. 
   }
    \label{fig:gainprofile}
\end{figure}

Figure~\ref{fig:gainprofile} shows the detector-noiseless and noisy speckle attenuation factor as the ratio of the  radial average of the median speckle subtracted (\gls{NMF} processed) image and the raw unprocessed speckle median image.
Figure~\ref{fig:gainprofile}(b), the noisy curve shows the $f_{\Delta I}$ at the \gls{IWA} is conservatively approximated by the assumed value of 0.25; equivalently, this is a post-processing gain of $4\times$.

The code to reproduce these figures is also available in the main publication repository \citep{douglas_exozodi_exosims_sensitivity_2020}.

\newpage
\appendix
HIPPARCOS \citep{esa_hipparcos_1997} identifiers of input catalog:
\input{HIP_names}
\end{document}

%% file: acronyms.tex
\newacronym{AU}{AU}{astronomical Unit [1.5e11 m]}  
\newacronym{pc}{pc}{parsec}
\newacronym{mas}{mas}{milliarcsecond}
\newacronym{nm}{nm}{nanometer}
\newacronym{CTE}{CTE}{coefficient of thermal expansion}
\newacronym{sqarc}{$as^2$}{square arcsecond}

\newacronym{smc}{SMC}{Small Magellanic Cloud}
\newacronym{lmc}{LMC}{Large Magellanic Cloud}
\newacronym{ism}{ISM}{interstellar medium}
\newacronym{mw}{MW}{Milky Way}
\newacronym{epseri}{$\epsilon$ Eri}{Epsilon Eridani}
\newacronym{EKB}{EKB}{Edgeworth-Kuiper Belt}

\newacronym{CFR}{CFR}{Complete Frequency Redistribution}

\newacronym{nasa}{NASA}{National Aeronautics and Space Agency}
\newacronym{esa}{ESA}{European Space Agency}
\newacronym{omi}{OMI}{\textit{Optical Mechanics Inc.}}
\newacronym{gsfc}{GSFC}{\gls{nasa} Goddard Space Flight Center}
\newacronym{stsci}{STScI}{Space Telescope Science Institute}
\newacronym{nsroc}{NSROC}{\gls{nasa} Sounding Rocket Operations Contract}
\newacronym{wff}{WFF}{\gls{nasa} Wallops Flight Facility}
\newacronym{wsmr}{WSMR}{White Sands Missile Range}

\newacronym{irac}{IRAC}{Infrared Array Camera}
\newacronym[plural=CCDs, firstplural=charge-coupled devices (CCDs)]{ccd}{CCD}{charge-coupled device}
\newacronym[plural=EMCCDs, firstplural=electron multiplying charge-coupled devices (EMCCDs)]{EMCCD}{EMCCD}{electron multiplying charge-coupled device}

\newacronym{DM}{DM}{Deformable Mirror}
\newacronym{MCP}{MCP}{ Microchannel Plate }
\newacronym{ipc}{IPC}{Image Proportional Counter}
\newacronym{cots}{COTS}{Commercial Off-The-Shelf}
\newacronym{ISR}{ISR}{incoherent scatter radar}
\newacronym{atcamera}{AT}{angle tracker}
\newacronym{MEMS}{MEMS}{microelectromechanical systems}
\newacronym{QE}{QE}{quantum efficiency}
\newacronym{RTD}{RTD}{Resistance Temperature Detector}
\newacronym{PID}{PID}{Proportional-Integral-Derivative}
\newacronym{PRNU}{PRNU}{photo response non-uniformity}
\newacronym{DSNU}{PRNU}{dark signal non-uniformity}
\newacronym{CMOS}{CMOS}{complementary metal–oxide–semiconductor}
\newacronym{TRL}{TRL}{technology readiness level}
\newacronym{swap}{SWaP}{Size, Weight, and Power}
\newacronym{ConOps}{ConOps}{concept of operations}
\newacronym{NRE}{NRE}{non-recurring engineering}
\newacronym{CBE}{CBE}{current best estimate}

\newacronym{FOV}{FOV}{field-of-view}
\newacronym{NIR}{NIR}{near-infrared}
\newacronym{PV}{PV}{Peak-to-Valley}
\newacronym{MRF}{MRF}{Magnetorheological finishing}
\newacronym{AO}{AO}{Adaptive Optics}
\newacronym{TTP}{TTP}{tip, tilt, and piston}
\newacronym{FPS}{FPS}{fine pointing system}
\newacronym{SHWFS}{SHWFS}{Shack-Hartmann Wavefront Sensor}
\newacronym{OAP}{OAP}{off-axis parabola}
\newacronym{LGS}{LGS}{laser guide star}
\newacronym{WFCS}{WFCS}{wavefront control system}
\newacronym{OPD}{OPD}{optical path difference}
\newacronym{UA}{UA}{University of Arizona}
\newacronym{MEL}{MEL}{Master Equipment List}
\newacronym{GEO}{GEO}{low-earth orbit}
\newacronym{LEO}{LEO}{geosynchronous orbit}
\newacronym{EFC}{EFC}{electric-field conjugation}
\newacronym{LDFC}{LDFC}{linear dark field control}
\newacronym{DAC}{DAC}{digital-to-analog converter}
\newacronym{FEA}{FEA}{finite element analysis}

\newacronym{SiC}{SiC}{Silicon Carbide}
\newacronym{ESPA}{ESPA}{EELV Secondary Payload Adapter}
\newacronym{EEID}{EEID}{Earth-equivalent insolation distance}
\newacronym{LLOWFS}{LLOWFS}{Lyot Low-order wavefront sensor}
\newacronym{WFSC}{WFSC}{wavefront sensing and control}
\newacronym{STOP}{STOP}{Structural-Thermal-Optical-Performance}

\newacronym{resel}{resel}{resolution element}

\newacronym{acs}{ACS}{Attitude Control System}
\newacronym{orsa}{ORSA}{Ogive Recovery System Assembly}
\newacronym{gse}{GSE}{Ground Station Equipment}
\newacronym{FSM}{FSM}{Fast Steering Mirror}

\newacronym{WFS}{WFS}{wavefront sensor}
\newacronym{LSI}{LSI}{Lateral Shearing Interferometer}
\newacronym{VVC}{VVC}{Vector Vortex Coronagraph}
\newacronym{VNC}{VNC}{Visible Nulling Coronagraph}
\newacronym{CGI}{Coronagraph}{Coronagraph}
\newacronym{IWA}{IWA}{Inner Working Angle}
\newacronym{OWA}{OWA}{Outer Working Angle}
\newacronym{NPZT}{N-PZT}{Nuller piezoelectric transducer}
\newacronym{ZWFS}{ZWFS}{Zernike wavefront sensor}
\newacronym{SPC}{SPC}{Shaped Pupil Coronagraph}
\newacronym{HLC}{HLC}{Hybrid-Lyot Coronagraph}
\newacronym{ADI}{ADI}{angular differential imaging}
\newacronym{RDI}{RDI}{reference differential imaging}

\newacronym{HST}{HST}{Hubble Space Telescope}
\newacronym{GPS}{GPS}{Global Positioning System}
\newacronym{ISS}{ISS}{International Space Station}
\newacronym[description=Advanced CCD Imaging Spectrometer]{acis}{ACIS}{Advanced \gls{ccd} Imaging Spectrometer}
\newacronym{stis}{STIS}{\textit{Space Telescope Imaging Spectrograph}}
\newacronym{mcp}{MCP}{Microchannel Plate}
\newacronym{jwst}{JWST}{$\textit{James Webb Space Telescope}$}
\newacronym{fuse}{FUSE}{$\textit{FUSE}$}
\newacronym{galex}{GALEX}{$\textit{Galaxy Evolution Explorer}$}
\newacronym{spitzer}{Spitzer}{$\textit{Spitzer Space Telescope}$}
\newacronym{mips}{MIPS}{Multiband Imaging Photometer for \gls{spitzer}}
\newacronym{gissmo}{GISSMO}{Gas Ionization Solar Spectral Monitor}
\newacronym{iue}{IUE}{International Ultraviolet Explorer}
\newacronym{spinr}{SPINR}{$\textit{Spectrograph for Photometric Imaging with Numeric Reconstruction}$}
\newacronym{imager}{IMAGER}{$\textit{Interstellar Medium Absorption Gradient Experiment Rocket}$}
\newacronym{TPF-C}{TPF-C}{Terrestrial Planet Finder Coronagraph}
\newacronym{RAIDS}{RAIDS}{Atmospheric and Ionospheric Detection System }
\newacronym{mama}{MAMA}{Multi-Anode Microchannel Array}
\newacronym{ATLAST}{ATLAST}{Advanced Technology Large Aperture Space Telescope}
\newacronym{PICTURE}{PICTURE}{Planet Imaging Concept Testbed Using a Rocket Experiment}
\newacronym{LITES}{LITES}{Limb-imaging Ionospheric and Thermospheric
Extreme-ultraviolet Spectrograph}
\newacronym{LBT}{LBT}{Large Binocular Telescope}
\newacronym{LBTI}{LBTI}{Large Binocular Telescope Interferometer}
\newacronym{KIN}{KIN}{Keck Interferometer Nuller}
\newacronym{SHARPI}{SHARPI}{Solar High-Angular Resolution Photometric Imager}
\newacronym{IRAS}{IRAS}{Infrared Astronomical Satellite}
\newacronym{HARPS}{HARPS}{High Accuracy Radial velocity Planetary}
\newacronym{hstSTIS}{STIS}{Space Telescope Imaging Spectrograph}
\newacronym{spitzerIRAC}{IRAC}{Infrared Array Camera}
\newacronym{spitzerMIPS}{MIPS}{Multiband Imaging Photometer for Spitzer}
\newacronym{spitzerIRS}{IRS}{Infrared Spectrograph}
\newacronym{CHARA}{CHARA}{Center for High Angular Resolution Astronomy}
\newacronym{wfirst-afta}{WFIRST-AFTA}{Wide-Field InfraRed Survey
Telescope-Astrophysics Focused Telescope Assets}
\newacronym{GPI}{GPI}{Gemini Planet Imager}
\newacronym{WFIRST}{WFIRST}{Wide-Field InfraRed Survey Telescope}
\newacronym{HabEx}{HabEx}{Habitable Exoplanet Observatory Mission Concept}
\newacronym{LUVOIR}{LUVOIR}{Large UV/Optical/Infrared Surveyor}
\newacronym{FGS}{FGS}{Fine Guidance Sensor}
\newacronym{STIS}{STIS}{Space Telescope Imaging Spectrograph}
\newacronym{MGHPCC}{MGHPCC}{Massachusetts Green High Performance
Computing Center}
\newacronym{WISE}{WISE}{Wide-field Infrared Survey Explorer}
\newacronym{ALMA}{ALMA}{Atacama Large Millimeter Array}
\newacronym{GRAIL}{GRAIL}{Gravity Recovery and Interior Laboratory}
\newacronym{jwstNIRCam}{NIRCam}{near-\gls{IR}-camera}
\newacronym{jwstMIRI}{MIRI}{Mid-Infrared Instrument}

\newacronym{AURIC}{AURIC}{The Atmospheric Ultraviolet Radiance Integrated Code} 
\newacronym{FFT}{FFT}{Fast Fourier Transform  }
\newacronym{MODTRAN}{MODTRAN   }{ MODerate resolution atmospheric TRANsmission }
\newacronym{idl}{IDL}{$\textit {Interactive Data Language}$}
\newacronym[sort=NED,description=NASA/IPAC Extragalactic Database]{ned}{NED}{\gls{nasa}/\gls{ipac} Extragalactic Database}
\newacronym{iraf}{IRAF}{Image Reduction and Analysis Facility}
\newacronym{wcs}{WCS}{World Coordinate System}
\newacronym{pegase}{PEGASE}{$\textit{Projet d'Etude des GAlaxies par Synthese Evolutive}$}
\newacronym{dirty}{DIRTY}{$\textit{DustI Radiative Transfer, Yeah!}$}
\newacronym{CUDA}{CUDA}{Compute Unified Device Architecture}
\newacronym{KLIP}{KLIP}{Karhunen-Lo\`{e}ve Image Processing}

\newacronym{MSIS}{MSIS}{Mass Spectrometer Incoherent Scatter Radar}
\newacronym{nmf2}{$N_m$}{F2-Region Peak density}
\newacronym{hmf2}{$h_m$}{F2-Region Peak height}
\newacronym{H}{$H$}{F2-Region Scale Height}

\newacronym{isr}{ISR}{Incoherent Scatter Radar}
\newacronym[description=TLA Within Another Acronym]{twaa}{TWAA}{\gls{tla} Within Another Acronym}
\newacronym[plural=SNe, firstplural=Supernovae (SNe)]{sn}{SN}{Supernova}
\newacronym{EUV}{EUV}{Extreme-Ultraviolet }
\newacronym{EUVS}{EUVS}{\gls{EUV} Spectrograph}
\newacronym{F2}{F2}{Ionospheric Chapman F Layer }
\newacronym{F10.7}{F10.7}{ 10.7 cm radio flux [10$^{-22}$ W m$^{-2}$ Hz$^{-1}$]  }
\newacronym{FUV}{FUV}{far-ultraviolet }
\newacronym{IR}{IR}{infrared}
\newacronym{MUV}{MUV}{mid-ultraviolet }
\newacronym{NUV}{NUV}{near-ultraviolet }
\newacronym{O$^+$}{O$^+$}{Singly Ionized Oxygen  Atom }
\newacronym{OI}{OI}{Neutral Atomic Oxygen Spectroscopic State }
\newacronym{OII}{OII}{Singly Ionized Atomic Oxygen Spectroscopic State }
\newacronym{PSF}{PSF}{point spread function}
\newacronym{$R_E$}{$R_E$}{Earth radii [$\approx$ 6400 km]  }
\newacronym{RV}{RV}{radial velocity}
\newacronym{UV}{UV}{ultraviolet }
\newacronym{WFE}{WFE}{wavefront error}
\newacronym{sed}{SED}{spectral energy distribution}
\newacronym{nir}{NIR}{near-infrared}
\newacronym{mir}{MIR}{mid-infrared}
\newacronym{ir}{IR}{infrared}
\newacronym{uv}{UV}{ultraviolet}
\newacronym[plural=PAHs, firstplural=Polycyclic Aromatic Hydrocarbons (PAHs)]{pah}{PAH}{Polycyclic Aromatic Hydrocarbon}
\newacronym{obsid}{OBSID}{Observation Identification}
\newacronym{SZA}{SZA}{Solar Zenith Angle}
\newacronym{TLE}{TLE}{Two Line Element set}
\newacronym{DOF}{DOF}{degrees-of-freedom}
\newacronym{PZT}{PZT}{lead zirconate titanate}
\newacronym{ADCS}{ADCS}{attitude determination and control system}
\newacronym{COTS}{COTS}{commercial off-the-shelf}
\newacronym{CDH}{C$\&$DH}{command and data handling}
\newacronym{EPS}{EPS}{electrical power system}

\newacronym{PCA}{PCA}{principal component analysis}
\newacronym{fwhm}{FWHM}{full-width-half maximum}
\newacronym{RMS}{RMS}{root mean squared}
\newacronym{RMSE}{RMSE}{root mean squared error}
\newacronym{MCMC}{MCMC}{Marcov chain Monte Carlo}
\newacronym{DIT}{DIT}{discrete inverse theory}
\newacronym{SNR}{SNR}{signal-to-noise ratio}
\newacronym{PSD}{PSD}{power spectral density}
\newacronym{NMF}{NMF}{non-negative matrix factorization}

%% file: conf_tab_values2_34alphaHLC-565-Disk.tex
27& 51 & 43 - 58 \\
9& 31 & 23 - 39 \\
3& 13 & 7 - 20 \\
1& 4 & 1 - 8 \\

%% file: HIP_names.tex
37279,
97649,
113368,
57632,
67927,
2021,
102422,
22449,
17378,
8102,
95501,
99240,
3821,
16537,
98036,
57757,
27072,
28103,
109176,
78072,
27321,
14632,
50954,
70497,
59199,
7513,
12777,
116771,
102485,
15510,
92043,
1599,
64394,
112447,
61317,
105858,
17651,
108870,
67153,
16852,
19849,
61174,
77257,
12843,
71284,
96100,
34834,
77760,
46509,
24813,
64924,
76829,
23693,
16245,
104214,
39903,
15457,
64408,
86486,
4151,
10644,
51459,
57443,
65721,
86736,
80686,
82860,
910,
47592,
5862,
109422,
73996,
29271,
53721,
25278,
97295,
48113,
95447,
88745,
86796,
7981,
29800,
25110,
97675,
40843,
45333,
104217,
64792,
3909,
15371,
56997,
114622,
32480,
47080,
49081,
80337,
73184,
84862,
78459,
12653,
18859,
58576,
32439,
89042,
79672,
22263,
8362,
15330,
7978,
99825,
3765,
35136,
107649,
12114,
38908,
42438,
105090,
81300,
75181,
98767,
26394,
3093,
49908,
84478,
43587,
3583,
23311,
56452,
72848,
40693,
13402,
100017,
114046,
54035,
27435,
544,
113283,
26779,
1475,
68184,
32984,
88972,
10798,
86400,
41926,
57939,
85235,
42808,
25878.